\input{psfig}
\documentstyle[prl,aps,twocolumn]{revtex}
\begin{document}
\draft
\def\simlt{\stackrel{<}{{}_\sim}}
\def\simgt{\stackrel{>}{{}_\sim}}
\date{\today}
\narrowtext
\pagestyle{empty}

\noindent {\bf Comment on ``Nongaussian Isocurvature Perturbations from
Inflation''}\\

In a recent Letter \cite{Lin} a hybrid inflationary model with 
nongaussian density perturbations and a ``blue'' spectral index 
$n \ge 1$ was presented. The purpose of this comment is to point out 
that this model can be considered as a particular realization of the 
general framework for the spectrum and statistics of the CMBR we proposed 
in \cite{sky}, based on the hypothesis of conformal invariance. 
The implication of this hypothesis is that density perturbations are
nongaussian with a spectral index $n\ge 1$. 

In \cite{Lin} the spectrum of density fluctuations is generated by a
free massive scalar field propagating in (approximately) de Sitter
background. The main contribution to the energy
density in this model comes from the modes with (comoving) wavelengths larger
than the horizon scale $H^{-1}$. In addition, the mean value of the
scalar field {\em vanishes} as a result of averaging over
sufficiently large scales, and the density perturbations
are quadratic in the scalar field fluctuations. Hence the
mechanism for generating the density perturbations proposed
in \cite{Lin} is not of the kind usually considered in 
inflationary models where a scalar inflaton field takes non-zero
values. A necessary consequence of this non-standard mechanism is
nongaussian statistics for the density perturbations and the CMBR.

These consequences of the model proposed in \cite{Lin} are
a specific realization of the general conformal invariance
considerations of our Letter. To see this we begin with the 
observation that a massive scalar field 
in de Sitter background exhibits scaling behavior for distances larger 
than the horizon. Indeed, its two--point function falls off with a power law
\cite{prop,ratra}:
\begin{eqnarray}
&&<\Phi(x)\Phi(x')>\sim {d(x,x')}^{-2\gamma} \label{gamma}\\
&&\gamma = {3\over 2} - \sqrt{{9\over 4} - {m^2\over H^2}} 
\approx {m^2\over 3H^2}\ ,
\end{eqnarray}
for $m \ll H$ and $d \gg H^{-1}$, where $m$ is the mass of $\Phi$ 
and $d$ is the geodesic distance. In spatially flat coordinates
\begin{equation}
d^2(x,x') = e^{H(t+t')}
\left[H^{-2}(e^{-Ht} - e^{-Ht'})^2 - ({\bf x}
- {\bf x'})^2\right].
\label{sigma}\end{equation}
Equation (\ref{gamma}) is just the definition of a conformal
field with dimension $\gamma$. Since the energy momentum tensor
is a quadratic form in $\Phi$ involving two derivatives, it follows 
that the scaling dimension
$\Delta$ of density perturbations due to the scalar field $\Phi$ is
$\Delta = 2\gamma + 2$, in the notation of \cite{sky}. Hence 
the spectral index 
\begin{equation}
n=2\Delta - 3 = 1 + 4\gamma \ .
\end{equation} Thus for $m^2 << H^2$ the general formula of
\cite {sky} reproduces the result
of \cite{Lin}, namely, $n\sim {1 + {4m^2\over 3H^2}}$.

{}The behavior of the two--point function (\ref{gamma}) 
at large distances is a reflection of the fact that there
can be no spontaneous symmetry breaking in de Sitter space \cite{ratra},
a situation closely analogous to the absence of spontaneous
breaking of global symmetries in two dimensions. It follows
that the mean value of the scalar $\Phi$ vanishes when averaging over
distances $d>d_0$, where $2\gamma \ln(H d_0) = {\cal O}(1)$,
or for comoving wavenumbers $k<<k_0$ with 
$k_0\sim H\exp(-{3H^2\over 2m^2})$ as pointed out in \cite{Lin}. 
It is this vanishing of the expectation value of the scalar field
which is responsible for the fact that one must go to quadratic
order in the fluctuations in $\Phi$ in order to obtain a non-vanishing
result for the density perturbations in the model of \cite{Lin}.
This implies necessarily the existence of higher point correlations
in the statistics of the density perturbations, and hence the CMBR.
These higher point correlations are determined by conformal
invariance as in \cite{sky}, since $\Phi$ 
scales like a conformal field at very large distances.

Let us note the close analogy between this vanishing of the
scalar field $\Phi$ when averaged over large distance scales
in the model of \cite{Lin} and the vanishing of the
average value of the Ricci scalar due to the quantum 
fluctuations of the conformal factor of the spacetime metric 
at large distance scales \cite{amm} in the model of \cite{sky}. 
It is this mechanism which gives rise to conformal invariance
and nongaussian statistics in the latter model.
The coefficient of the trace anomaly $Q^2$ which gives rise 
to these fluctuations leads to an anomalous dimension of the 
scaling behavior and a ``blue'' spectrum for the CMBR with 
$4/Q^2$ playing essentially the same role as $4m^2/3H^2$. 
This parameter is naturally of order one in the conformal 
factor theory, without the need to introduce a 
new supersymmetric sector. 

I. A. and E. M. acknowledge the hospitality of the Dept. of
Physics and Astronomy, Univ. of S. Carolina.

\vspace{.4cm}

\noindent {\small Ignatios Antoniadis$^1$, Pawel O. Mazur$^2$,
and Emil Mottola$^3$

$^1$ Centre de Physique Th\'eorique,

Ecole Polytechnique, 91128 Palaiseau, France and

Theory Division, CERN, 1211 Geneva 23, Switzerland

$^2$ Department of Physics and Astronomy,

University of South Carolina, Columbia, SC 29208

$^3$ Theoretical Division, Los Alamos National Laboratory, 

MS B285, Los Alamos, NM 87545}

\end{document}